\documentclass[epj]{svjour}
\usepackage{latexsym}
\usepackage{graphics}
\usepackage[utf8]{inputenc}
\usepackage{graphics,graphicx}
\usepackage{amsmath,amssymb,amsfonts,latexsym,cancel}

\begin{document}

\title{Stable relativistic polytropic objects with cosmological constant}
\author{Jos\'e D. V. Arba\~nil\inst{1}
        \and
        Pedro H. R. S. Moraes\inst{2,3} 
        }

\institute{Departamento de Ciencias, Universidad Privada del Norte, Avenida el Sol 461 San Juan de Lurigancho, 15434 Lima,  Peru \mail{jose.arbanil@upn.pe}
          \and
Departamento de F\'isica, Instituto Tecnol\'ogico de Aeron\'autica, Centro T\'ecnico Aeroespacial, S\~ao Jos\'e dos Cam\-pos 12228-900, S\~ao Paulo, Brazil \and
Instituto de Astronomia, Geof\'isica e Ci\^encias Atmosf\'ericas, Universidade de S\~ao Paulo, Rua do Mat\~ao 1226 Cidade Universitária, São Paulo 05508-090, SP, Brazil
}

\date{Received: date / Accepted: date}

\abstract{
The effects of the cosmological constant on the static equilibrium configurations and stability against small radial perturbations of relativistic polytropic spheres are investigated. This study numerically solves the hydrostatic equilibrium equation and the radial stability equation, both of which are modified from their standard form to introduce the cosmological constant. For the fluid, we consider a pressure $p$ and an energy density $\rho$, which are connected through the equation of state $p=\kappa\delta^{\Gamma}$ with $\delta=\rho-p/(\Gamma-1)$, where $\kappa$, $\Gamma$ and $\delta$ represent the polytropic constant, adiabatic index and rest mass density of the fluid, respectively. The dependencies of the mass, radius and eigenfrequency of oscillations on both the cosmological constant and the adiabatic index are analyzed. For ranges of both the central rest mass density $\delta_c$ and the adiabatic index $\Gamma$, we show that the stars have a larger (lower) mass and radius and a diminished (enhanced) stability when the cosmological constant $\Lambda>0$ ($\Lambda<0$) is increased (decreased). In addition, in a sequence of compact objects with fixed $\Gamma$ and $\Lambda$, the regions constructed by stable and unstable static equilibrium configurations are recognized by the conditions $dM/d\delta_c>0$ and $dM/d\delta_c<0$, respectively.
\PACS{
      {}{Compact objects}   \and
      {}{Cosmological constant}
     } 
}
\maketitle

\section{Introduction}\label{sec-introd}

In 1917, the cosmological constant $\Lambda$ was introduced by A. Einstein in his general relativity field equations, yielding \cite{einstein/1917}
\begin{equation}\label{field_eq}
    G_{\mu\nu}+\Lambda g_{\mu\nu}=8\pi T_{\mu\nu}.
\end{equation}
In Eq.~(\ref{field_eq}), $G_{\mu\nu}$ is the Einstein tensor, $g_{\mu\nu}$ is the metric tensor, $T_{\mu\nu}$ is the energy-momentum tensor and $c=1=G$ units are assumed. 

It is common to see $\Lambda$ moved to the {\it rhs} of Equation (\ref{field_eq}) as
\begin{equation}\label{i2}
    G_{\mu\nu}=8\pi(T_{\mu\nu}+T_{\mu\nu}^{\rm vac}),
\end{equation}
with 
\begin{equation}\label{i3}
    T_{\mu\nu}^{\rm vac}=-\frac{\Lambda}{8\pi}g_{\mu\nu},
\end{equation}
and in this sense, $T_{\mu\nu}^{\rm vac}$ is the vacuum energy-momentum tensor. Note that even in the absence of matter ($T_{\mu\nu}=0$), the total energy-momentum tensor in Eq.~(\ref{i2}) is $\neq0$. Moreover, the effects of $T_{\mu\nu}^{\rm vac}$ are expected to be gravitational only; that is, the effects cannot be ``observed'' otherwise. In this particular regard, it is worth mentioning that a positive cosmological constant has, in fact, an ``anti-gravitational'' or repulsive effect. This will be revisited soon. 

Einstein`s intention with the presence of $\Lambda$ in Eq.~(\ref{field_eq}) was to predict a static, unchanging universe, which was the belief at that time. Thirteen years later, with E. Hubble's discovery of galaxies with receding velocities \cite{hubble/1929} and, consequently, an expanding universe, Einstein had to forfeit his idea, considering the cosmological constant, in his own words, ``the biggest blunder'' of his life.

Remarkably, the cosmological constant has not only been revived but also introduced unprecedented disarray into the scientific community. While the cosmological constant properly fits cosmological observations, which points to an accelerated expanding universe  \cite{riess/1998,planck_collaboration/2016,hinshaw/2013,perlmutter/1999}, the cosmological constant is haunted by ``the worst theoretical prediction in the history of physics'' \cite{hobson/2006}, namely, the ``cosmological constant problem'' \cite{weinberg/1989,padmanabhan/2003,carroll/1992,sahni/2000}. In other words, to fit the cosmological observations, the value of the cosmological constant must be many orders of magnitude smaller than the value predicted from unified theories of elementary particles. This issue is nothing but the ``the weight of the vacuum'' \cite{padmanabhan/2003}.

There have been some attempts to solve the cosmological constant problem. One possibility is to examine the observational consequences of a vacuum energy that decays in time \cite{freese/1987}, disguised in $\Lambda(t)$ models \cite{overduin/1998,berman/1991,pan/2018,basilakos/2009,sola/2015}. P. D. Mannheim has argued, on the other hand, that to solve the cosmological constant problem, one does not need to change or quench the energy content of the universe but rather only its effect on cosmic evolution \cite{mannheim/2001}. In this case, $\Lambda$ actually becomes as large as elementary particle physics suggests (recall \cite{weinberg/1989}). Other attempts can be seen, for instance, in unimodular gravity \cite{percacci/2018,padilla/2015,alvarez/2015,smolin/2009}, holographic dark energy  \cite{wang/2017,huang/2012} and extra-dimensional models \cite{arkani-hamed/2000,dvali/2003,gunther/2003,koyama/2008}. 

To avoid the cosmological constant problem, it is common to see some attempts in the literature to describe cosmic acceleration without a need for the cosmological constant \cite{hu/2007,amendola/2000,armendariz-picon/2001,cai/2016,cognola/2006} (see also References \cite{smsb/2018,saez-gomez/2016,ms/2016,clifton/2015}). Nevertheless, the cosmological constant (standard) model is the best and simplest way to fit cosmological data.

The effect of the cosmological constant is analyzed in many astrophysically motivated problems, for example, focusing on the properties of the radial trajectories of test particles \cite{stuchlik/1983} and of geometrically thin accretion disks \cite{stuchlik/2005}, exploring circular orbits \cite{stuchlik/1999} and investigating the equilibrium configuration of a perfect fluid orbiting Schwarzschild-de Sitter black holes \cite{stuchlik_slany/2000}. Moreover, the role of the cosmological constant in the galaxies neighboring the Milky Way has been tested; for instance, this topic has been explicitly investigated for both small and large Magellanic Clouds in the gravitational field of the Milky Way \cite{stuchlik/2011}.

In the present paper we wish to methodically investigate the role of a cosmological constant in the stellar equilibrium configuration and radial stability of compact astrophysical objects. Such objects have been profoundly studied in alternative gravity theories (as seen in, for instance, References \cite{pani/2011,momeni/2015,orellana/2013}, among many others). Most of the time, in addition to investigating the possibility of existence of these objects in such theories, these works are motivated by an attempt to predict the existence of massive pulsars, such as those reported in \cite{antoniadis/2013,demorest/2010,linares/2018}.

Compact objects in the presence of a cosmological constant can be seen in \cite{bordbar/2016,largani/2019,nayak/2015,liu/2019}. In \cite{bordbar/2016}, G.H. Bordbar et al. obtained a maximum mass of $\sim1.68M_\odot$ for a neutron star of $\Lambda\sim10^{-8}$\,km$^{-2}$ (a value significantly distinct from the one predicted via cosmological observations, namely $\Lambda\sim10^{-46}$\,km$^{-2}$, which we will revisit in our approach). As reported in \cite{largani/2019}, the typical values of $\Lambda$ that yield observable effects in the structure of a neutron star are $\sim10^{-2}$\,km$^{-2}$, which is probably a consequence of the stiff equation of state applied by these authors. In \cite{nayak/2015}, it was shown that the maximum mass and radius of a star increase with increasing $\Lambda$. For a quark-meson coupling model equation of state, it is possible to obtain a maximum mass similar to the recently reported massive pulsar in Reference \cite{antoniadis/2013}. In addition, the influence of the cosmological constant on the structure configuration of a compact object with a Soffer equation of state has also been investigated. For instance, in \cite{liu/2019}, the effect of $\Lambda$ on the physical properties of a white dwarf is analyzed; in this work, the authors find that such structure configurations are affected by a cosmological constant of $10^{-10}\,{\rm km}^{-2}$ (a lower value than the one required for a neutron star). It is important to mention that large values of the cosmological constant have been considered in other types of astrophysical situations. For example, in the case of accretion in primordial black holes during the very early universe, the cosmological constant can take values many orders of magnitude greater than $10^{-46}\,{\rm km}^{-2}$ (see \cite{stuchlik_slany/2000}). Thus, in the aforementioned works, motivated by the search for new equilibrium configurations or new phenomena responsible for structure equilibrium configurations, the authors consider a cosmological constant value larger than the one predicted by cosmological observations.

Our paper is organized as follows. The stellar structure and radial stability equations are derived in Section 2. We also present the equation of state we assume herein, namely, a relativistic polytropic equation of state. The stellar structure and radial stability equations are solved in Section 3, in which we present the numerical method employed to solve these equations. Our results are presented for different values of $\Lambda$, including $\Lambda=0$ for the sake of completeness. We show different profiles for the resulting stars that involve their mass, central rest mass density, polytropic index, fundamental-mode eigenfrequency and radius. Our conclusions are presented in Section 4.


\section{General relativistic formulation}\label{sec-basicequations}

\subsection{Stellar structure equations}

The perfect fluid inside the compact object considered herein is described by the stress-energy tensor, which can be expressed as
\begin{eqnarray}\label{tem}
T_{\mu\nu}=(p+\rho)U_{\mu}U_{\nu}+p\,g_{\mu\nu},
\end{eqnarray}
where $p$, $\rho$ and $U_{\mu}$ are the fluid pressure, the fluid energy density and its four-velocity, respectively.

With the aim of analyzing the properties of the fluid within the spherically symmetric compact object, we use a spacetime line element in Schwarzschild-like coordinates as follows:
\begin{equation}\label{metric}
ds^2=-e^{\nu}dt^2+e^{\lambda}dr^2+r^2d\theta^2+r^2\sin^2\theta d\phi^2,
\end{equation}
with the functions $\nu=\nu(r)$ and $\lambda=\lambda(r)$ being dependent on the radial coordinate only. 

For the assumed stress-momentum tensor, Eq.~(\ref{tem}), and line element, Eq.~(\ref{metric}), we find that the nonzero components of the Einstein's field equations, Eq.~(\ref{field_eq}), are
\begin{eqnarray}
&&\frac{dm}{dr}=4\pi\rho r^2,\label{eq_masa}\\
&&\frac{dp}{dr}=-(p+\rho)\left(4\pi rp+\frac{m}{r^2}-\frac{\Lambda r}{3}\right)e^{\lambda},\label{tov2}\\
&&\frac{d\nu}{dr}=-\frac{2}{(p+\rho)}\frac{dp}{dr},\label{eq_nu}
\end{eqnarray}
where
\begin{equation}\label{eq_lambda}
e^{\lambda}=\left(1-\frac{2m}{r}-\frac{\Lambda r^2}{3}\right)^{-1}. 
\end{equation}
In the equations above, $m$ represents the mass function within the radius $r$. Eqs.~(\ref{eq_masa})--(\ref{eq_lambda}) are known as the stellar structure equations. Eq.~(\ref{eq_masa}) is known as the Tolman-O\-ppen\-hei\-mer-Volkoff (TOV) or hydrostatic equilibrium equation \cite{tolman,oppievolkoff}, which is altered from its standard form to introduce the cosmological constant $\Lambda$ \cite{bohmer_harko2005}.

With the purpose of finding static equilibrium configurations, Eqs.~(\ref{eq_masa})--(\ref{eq_lambda}) are integrated along the radial coordinate $r$. The conditions at the center of the object, $r=0$, are:
\begin{equation}
m(0)=0,\;\;\rho(0)=\rho_c,\;\;p(0)=p_c\;\;{\rm and}\;\;\nu(0)=\nu_c.
\end{equation}
The surface of the sphere, $r\rightarrow R$, is found by the condition $p(R)\rightarrow0$. At this point, the interior solution is connected with the Schwarzschild-de Sitter vacuum exterior solution. This means that, at the surface of the object, the interior and exterior metric functions are connected as follows:
\begin{equation}
e^{\nu(R)}=\frac{1}{e^{\lambda(R)}}=1-\frac{2M}{R}-\frac{\Lambda R^2}{3},
\end{equation}
where $M$ represents the total mass of the object.

\subsection{Radial stability equations}

The radial stability equations are obtained by infinitesimally perturbing the fluid variables and potential metrics. This is accomplished by preserving the spherical symmetry of the background object. The perturbed quantities are inserted into the Einstein field equations and the stress-energy momentum tensor conservation while preserving only the first-order terms.

In \cite{chandrasekhar_rp,chandrasekhar_PRL}, Chandrasekhar reported the radial stability equation for the first time. Its solution supplies information on the eigenfrequency of oscillation $\omega$. It is well known that, to rearrange this equation into a more adequate form for a numerical solutions, the Chandrasekhar equation can be split into two first-order equations for the functions $\Delta r/r$ and $\Delta p$, with $\Delta r$ and $\Delta p$ being the relative radial displacement and Lagrangian perturbation of pressure, respectively (see, for example, \cite{vath_chanmugam1992,gondek1997,gondek1999}). 

The effects of the cosmological constant on the radial stability of both incompressible objects and polytropic spheres were investigated in \cite{stuchlik_proc/2005} by inserting the cosmological constant into the general relativistic field equations, thus generalizing the radial pulsation equation with the presence of $\Lambda$. To obtain more appropriate equations for numerical solutions, B\"ohmer and T. Harko \cite{bohmer_harko2005} presented these equations, for $\xi=\Delta r/r$, in the following form:
\begin{eqnarray}
&&\frac{d\xi}{dr}=\frac{\xi}{2}\frac{d\nu}{dr}-\frac{1}{r}\left(3\xi+\frac{\Delta p}{p \Gamma}\right),\label{ro1}\\
&&\frac{d\Delta p}{dr}= (p+\rho)\omega^2\xi re^{\lambda-\nu}+\left(\frac{d\nu}{dr}\right)^2\frac{(p+\rho)\xi r}{4}\nonumber\\
&&-4\xi\left(\frac{dp}{dr}\right)-(p+\rho)\left(8\pi p-\Lambda\right)\xi r e^{\lambda}-\left[\frac{1}{2}\frac{d\nu}{dr}+4\pi re^\lambda(p+\rho)\right]\Delta p,\label{ro2}
\end{eqnarray}
where $\Gamma=\left(1+\frac{\rho}{p}\right)\frac{dp}{d\rho}$ represents the adiabatic index. The functions $\xi$ and $\Delta p$ are considered to have a time dependence of the form $e^{i\omega t}$.

To analyze the stability against small radial perturbations, Eqs.~(\ref{ro1}) and (\ref{ro2}) are integrated from the center to the surface of the object. To achieve regularity in the center of the sphere, it is required that
\begin{equation}
\left[\Delta p\right]_{\rm center}=-3\left[\xi\Gamma p\right]_{\rm center}.
\end{equation}
At this point, for normalized eigenfunctions, we have $\xi(r=0)=1$. In turn, on the surface of the object $r=R$, it is found that
\begin{equation}\label{delta_p}
\left[\Delta p\right]_{\rm surface}=0.
\end{equation}

\subsection{Relativistic polytropic equation of state and speed of sound}

For the present work, we use the relativistic polytropic equation of state (EoS) \cite{tooper}. This EoS determines that the pressure $p$ and the energy density $\rho$ of the fluid obey the equalities
\begin{equation}\label{EoS2}
p=\kappa\delta^{\Gamma}\;\;\;\;{\rm and}\;\;\;\;\rho=\delta+p/(\Gamma-1),
\end{equation}
where $\kappa$ and $\delta$ are the polytropic constant and the rest mass density, respectively. Following \cite{raymalheirolemoszanchin,alz-poli-qbh,alz-2eos-qbh}, we consider the following polytropic constant:
\begin{equation}
\kappa=1.47518\times10^{-3}\left(1.78266\times10^{15}{\rm kg/m^3}\right)^{1-\Gamma}.
\end{equation}
It is important to highlight that, as mentioned in reference \cite{alz-2eos-qbh}, in equilibrium configurations where the fluid pressure is small relative to the energy density, EoS \eqref{EoS2} is similar to the nonrelativistic polytropic EoS $p=\kappa\rho^{\Gamma}$. This relation is used to investigate the influence of the cosmological constant on the spherical equilibrium configurations and radial stability of polytropes (see, for instance \cite{stuchlik_proc/2005,stuchlik/2016}). Moreover, this relation is used to address the extremely compact polytropes (with a high exponent $\Gamma$) \cite{stuchlik/2017,novotny/2017} and, if we extrapolate the polytropic exponent $\Gamma$ to infinity, to achieve an incompressible fluid configuration \cite{stuchlik/2000,bohmer/2004}. A discussion about the compactness of compact objects can be found in \cite{arbanil/2014,hod/2018}.

With the objective of checking where the causality limit may be violated, we need to analyze the speed of sound of a compressible fluid through the following relation:
\begin{equation}
c_s^2=\frac{dp}{d\rho}.
\end{equation}
For a relativistic polytropic fluid defined according to Eq.~\eqref{EoS2}, this relation yields
\begin{equation}
c_s^2=\frac{p\Gamma }{p+\rho}.
\end{equation}
For a given $\Gamma$, the maximum ratio $p/\delta$ that guarantees $c_s\leq1$ is
\begin{equation}
\frac{\Gamma(\Gamma-2)}{\Gamma-1}\leq\frac{\delta}{p},
\end{equation}
which indicates that the causality condition is not violated for $\Gamma$ in the range of $1\leq\Gamma\leq2$. Moreover, in order to assure that $c_s\leq1$ for $\Gamma>2$, for higher values of $\Gamma$, a relatively larger $\delta/p$ is required.

\section{Equilibrium and stability of relativistic polytropic spheres}

\subsection{Numerical method}

To investigate the static equilibrium configuration of spherically symmetric objects, we solve the stellar structure equations by means of the Runge-Kutta fourth-order method for a given $\Gamma$, $\delta_c$ and $\Lambda$. Then, the radial oscillation equations are solved through the shooting method. This process begins with the numerical integration of Eqs.~(\ref{ro1}) and (\ref{ro2}) for a test value of $\omega^2$. After each integration, the test value is corrected; this continues until condition (\ref{delta_p}) is achieved in the subsequent integration. The values of $\omega$ that satisfy this last condition are called the eigenfrequencies of oscillation. It is important to mention that our numerical algorithm reproduces the results presented in References ~\cite{gondek1999,benvenuto_horvath1991}.

With the purpose of numerically investigating the equilibrium and stability of relativistic polytropic objects for different adiabatic indexes, the normalization factors must be chosen carefully. A typical normalization factor considered for the solutions of the TOV equation and radial perturbation equation is $\delta_0=1.78266\times10^{15}{\rm kg/m^3}$, which is the same factor considered in \cite{alz-2eos-qbh}.

The results below are found considering different central rest mass densities $\delta_c$, adiabatic indexes $\Gamma$ and cosmological constants $\Lambda$. In Subsection \ref{EQ_RS_lambda_0}, we investigate the static equilibrium and radial stability of relativistic polytropic objects for the central rest mass densities $10^{13}\leq\delta_c\leq10^{20}{\rm kg/m^3}$ and the adiabatic indexes $5/3\leq\Gamma\leq9/3$. In Subsection \ref{EQ_RS_lambda_neq0}, since we are interested in evaluating the influences of the cosmological constant on some physical properties of very compact objects, following \cite{alz-2eos-qbh}, we consider $\delta_c=10\delta_0$. For this central rest mass density, the maximum adiabatic index that yields good results is  $\sim4$.

\subsection{Equilibrium configuration and radial stability of relativistic polytropic spheres with $\Lambda=0$}\label{EQ_RS_lambda_0}

\begin{figure}[ht]
\begin{center}
\includegraphics[width=0.49\linewidth]{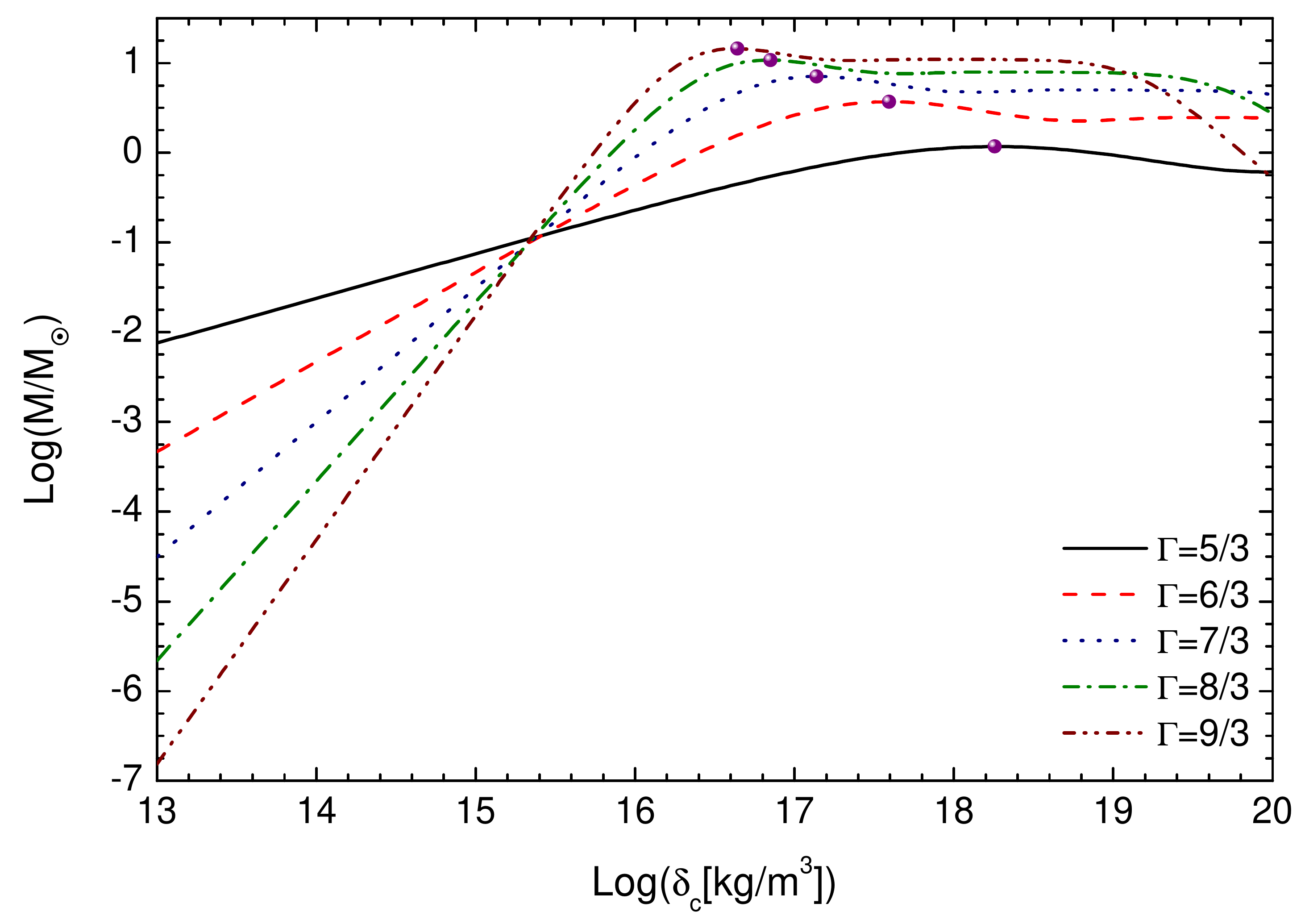}
\includegraphics[width=0.49\linewidth]{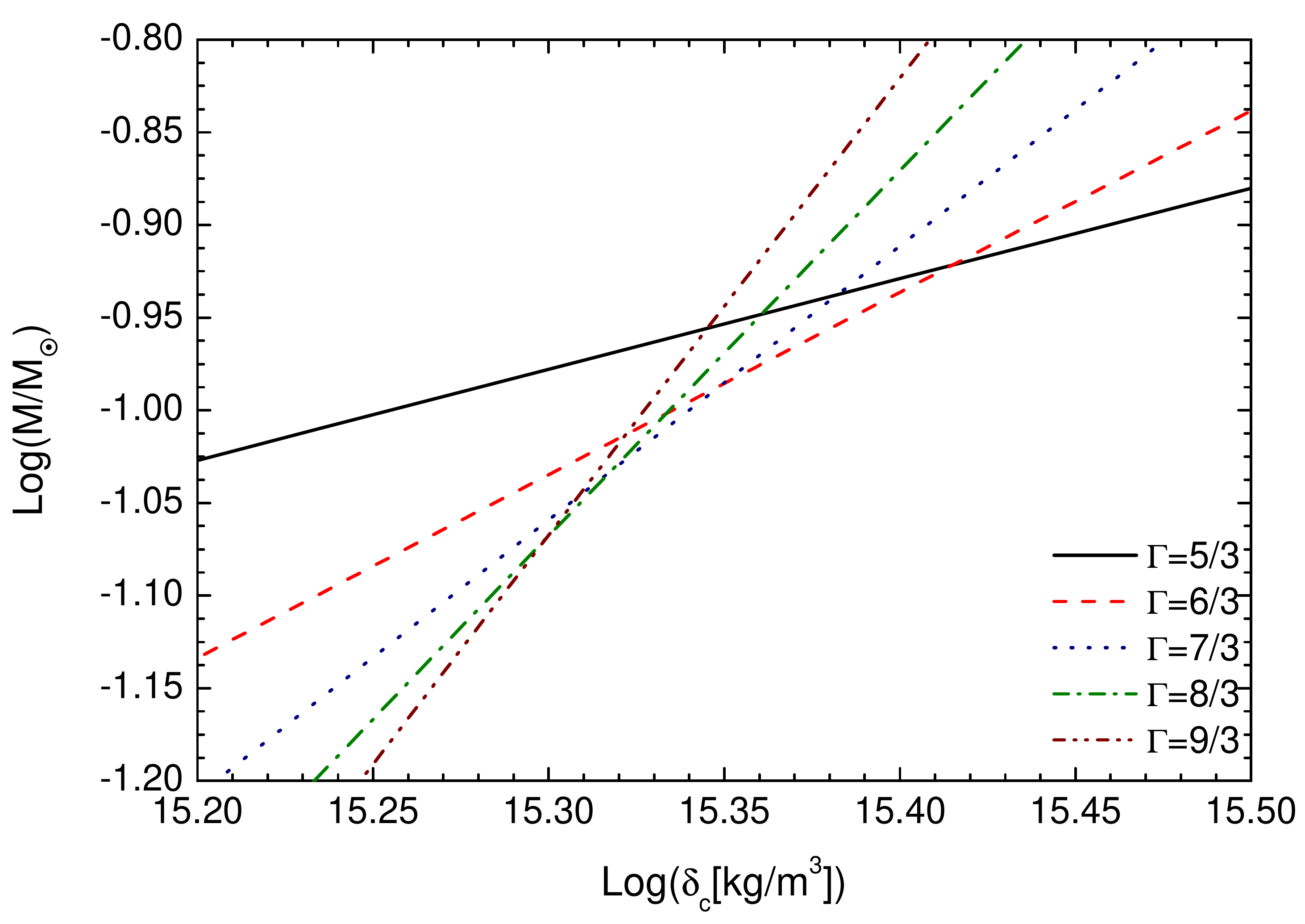}
\caption{Left: Mass as a function of the central rest mass density for some different values of $\Gamma$ and $\Lambda=0$. Right: Magnification of the region where the curves intersect. Note that the curves do not converge at the same point, as it might be deduced from the left panel.}
\label{Log_rhoc_Log_M}
\end{center}
\end{figure}

The behavior of the total mass with the central rest mass density is presented in Fig.~\ref{Log_rhoc_Log_M} for some different polytropic indexes $\Gamma$. The central rest mass density $\delta_c$ runs from $1.0\times10^{13}$ to $1.0\times10^{20}{\rm kg/m^3}$. The full circles on the curves mark the maximum mass points. In all curves, we can note the monotonic increase in the mass with the central rest mass density until it attains the maximum mass value at $\delta_c^*$. After this point, the mass decreases with the increase in $\delta_c$. 

The change in the total mass with the polytropic index is also observed in Fig.~\ref{Log_rhoc_Log_M}, where three regions are shown. In the first region, where $\delta_c\lesssim2.0\times10^{15}{\rm kg/m^3}$, the total mass decreases with an increase in the polytropic index. In the second region, where $\delta_c\gtrsim2.6\times10^{15}{\rm kg/m^3}$, the total mass grows with an increase in $\Gamma$. The third region, where $2.0\times 10^{15}\lesssim\delta c\lesssim 2.6\times 10^{15}{\rm kg/m^3}$ we see how the dependence of the mass on the polytropic index changes from the first to the second region. 

\begin{figure}[ht]
\begin{center}
\includegraphics[width=0.49\linewidth]{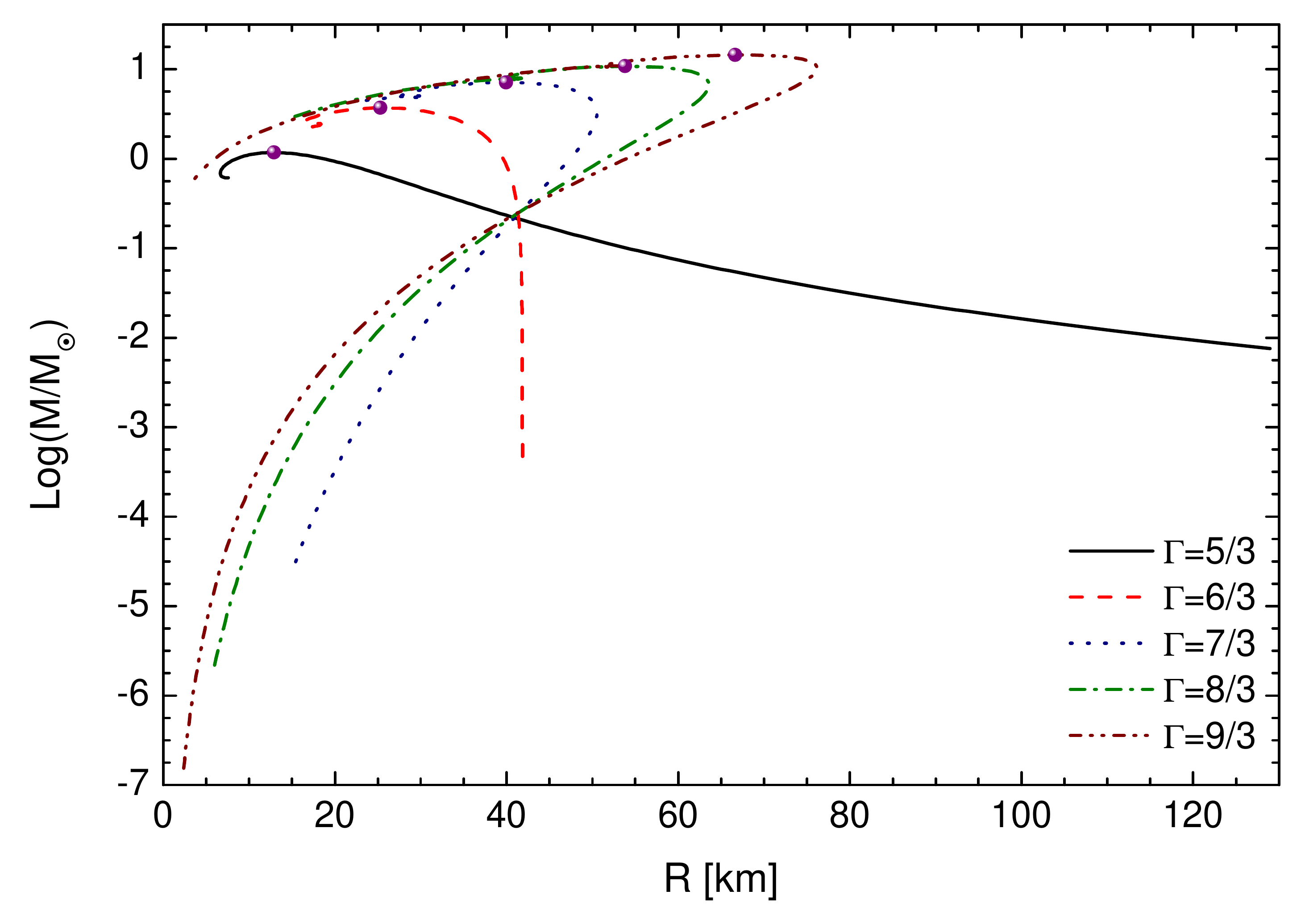}
\caption{Mass as a function of the total radius for five different polytropic indexes $\Gamma$ and $\Lambda=0$.}
\label{R_Log_M_gamma}
\end{center}
\end{figure}

In Fig.~\ref{R_Log_M_gamma}, we show the mass against the total radius for five different polytropic exponents. As in Fig.~\ref{Log_rhoc_Log_M}, the regarded central rest mass densities are in the range from $1.0\times10^{13}$ to $1.0\times10^{20}[\rm kg/m^3]$. Although these curves are very similar to those  derived by using the nonrelativistic polytropic EoS \cite{alz-poli-qbh}, the relativistic and nonrelativistic results differ in the large energy density (rest mass density) regime \cite{alz-2eos-qbh}. 

It is important to note that the inclination of the curve $M\times R$ becomes clockwise with an increase of $\Gamma$ and becomes approximately vertical at $\Gamma\approx2.0$. From Figs.~\ref{Log_rhoc_Log_M} and \ref{R_Log_M_gamma}, a large polytropic exponent corresponds to a constant rest mass density, $\delta_c=\delta_{0}$, giving rise to the relation $M\sim 4\pi\delta_0 R^3/3$. From this, it can be understood that the mass grows linearly with the total radius cubed.

\begin{figure}[ht]
\begin{center}
\includegraphics[width=0.49\linewidth]{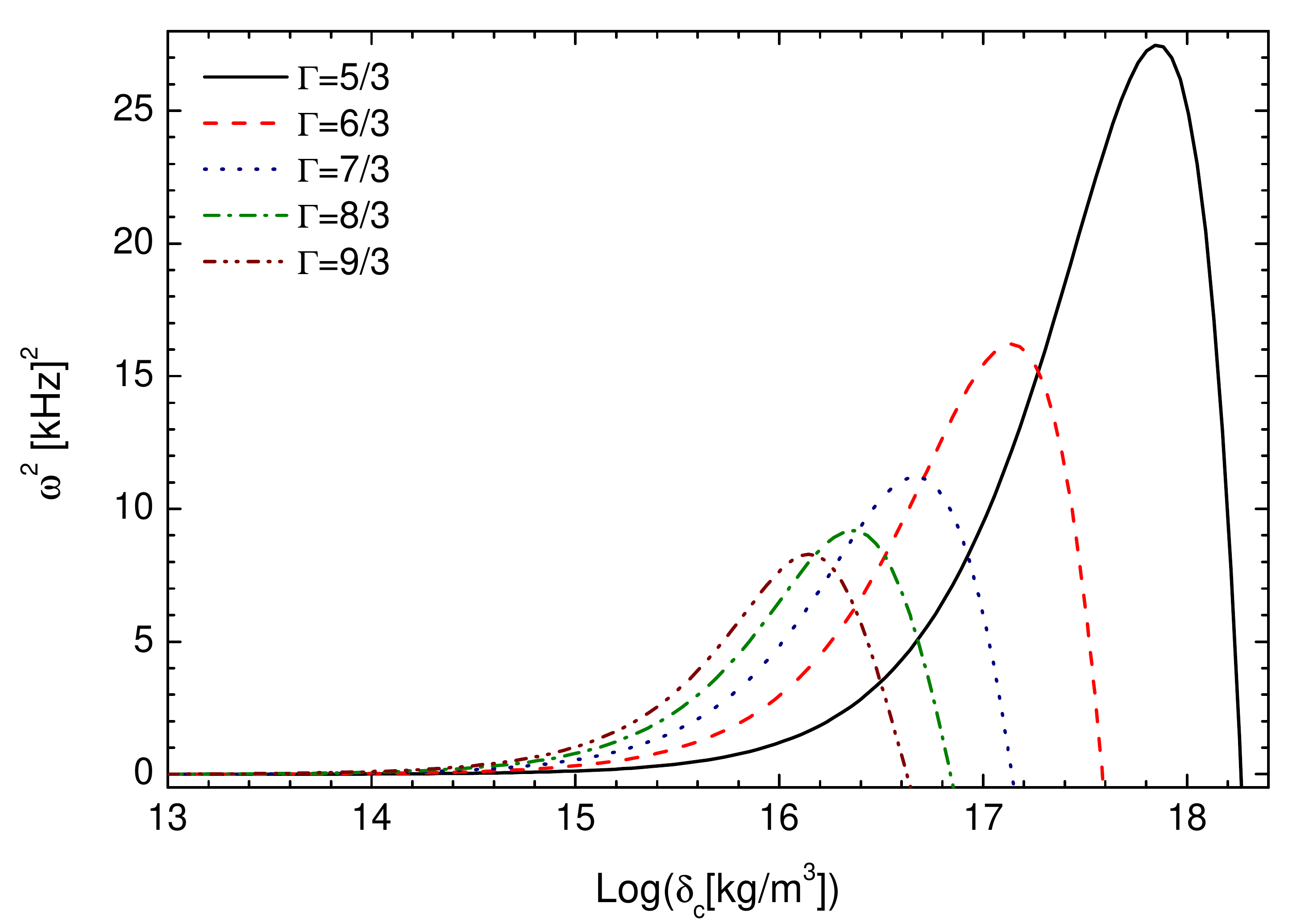}
\caption{Fundamental-mode eigenfrequency of oscillation squared against the central rest mass density for some values of $\Gamma$ and $\Lambda=0$.}
\label{Log_rhoc_w2_gamma}
\end{center}
\end{figure}

The behavior of the fundamental-mode eigenfrequency of oscillation squared, $\omega^2$, with the central rest mass density is plotted in Fig.~\ref{Log_rhoc_w2_gamma} for five polytropic indexes $\Gamma$. This figure considers stable relativistic polytropic spheres against small radial perturbations. In all cases analyzed, the functions $\omega^2(\delta_c)$ obey a Gaussian distribution. This function attains $\omega^2=0$ at the maximum total mass values. From this, it can be understood that, independent of the polytropic index used, the maximum mass point indicates the onset of instability (see Fig.~\ref{Log_rhoc_Log_M}). Thus, in a sequence of polytropic compact objects with the same polytropic index, the necessary and sufficient conditions required to identify regions composed of stable and unstable stars against small radial perturbations are $dM/d\delta_{c}>0$ and $dM/d\delta_{c}< 0$, respectively. This method is similar to the turning-point method for the axisymmetric stability of rotating relativistic stars \cite{friedman1988,sorkin1982} (see also \cite{takami2011}), for the radial stability of charged strange stars \cite{arbanil_malheiro}, for anisotropic strange stars \cite{arbanil_malheiro2016} and for strange stars in $d$ dimensions \cite{arbanil_malheiroPRD2019}. In these works, in a sequence of equilibrium configurations with the angular momentum, total electric charge, anisotropy at the star surface and the spacetime dimension fixed, the maximum total mass and the zero eigenfrequency of oscillation are derived for the same central energy density. 

On the other hand, in Fig. \ref{Log_rhoc_w2_gamma}, we can also note that a relatively low $\Gamma$ admits stable equilibrium configurations with larger $\delta_c$. In addition, for low central rest mass densities ($\lesssim10^{16}\,{\rm kg/m^3}$), the radial stability of relativistic polytropic spheres increases with $\Gamma$.

\subsection{Equilibrium configuration and radial stability of relativistic polytropic spheres with $\Lambda\neq0$}\label{EQ_RS_lambda_neq0}

\subsubsection{Equilibrium configuration and radial stability as a function of $\delta_c$}

In all relativistic polytropic structure configurations analyzed, we find that the effects of the cosmological constant are more visible at relatively low central rest mass densities. In these cases, the hydrostatic structure configurations are affected by the cosmological constants of $\sim10^{-5}{\rm km}^{-2}$ ($-10^{-5}{\rm km}^{-2}$), which are lower (larger) than those used for strange stars with values of $\Lambda\sim10^{-3}{\rm km}^{-2}$ ($-10^{-3}{\rm km}^{-2}$) \cite{zubairi2015}. Herein, we infer that the effect of the cosmological constant is more notable in objects that are relatively less compact. This finding is consistent with the results reported in the literature, where the cosmological constant values needed to affect the physical properties of white dwarfs \cite{liu/2019} are lower than those used in denser objects such as neutron stars \cite{bordbar/2016}. 

Furthermore, in a sequence of equilibrium configurations with the adiabatic index and cosmological constant fixed, the maximum mass peak matches with the zero eigenfrequency of oscillation. This indicates that stable and unstable equilibrium regions are determined by the relations $dM/d\delta_c>0$ and $dM/d\delta_c<0$, respectively.

\subsubsection{Equilibrium configuration and radial stability as a function of $\Gamma$}

\begin{figure}[ht]
\begin{center}
\includegraphics[width=0.49\linewidth]{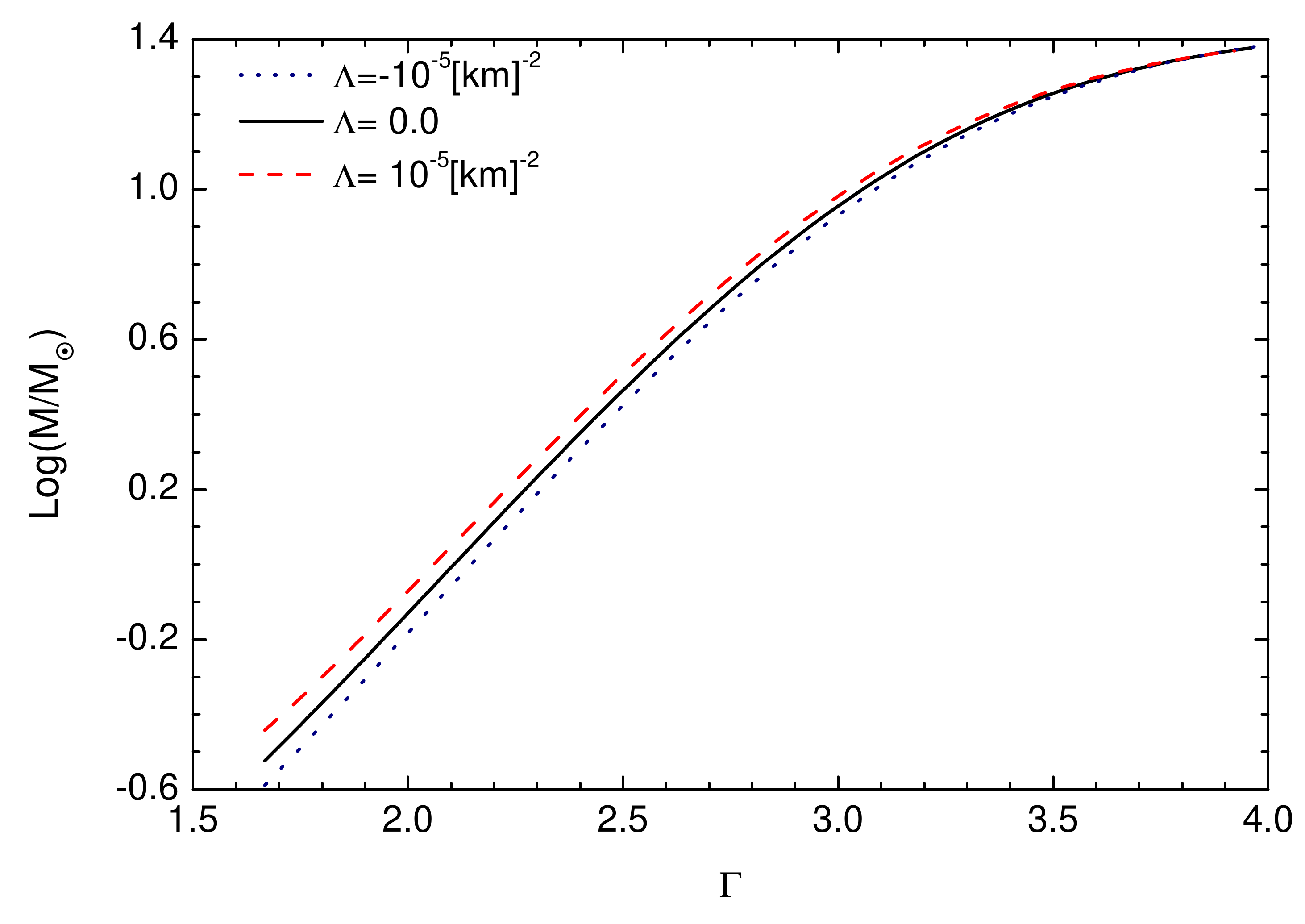}
\caption{Mass as a function of the adiabatic index for some cosmological constant values.}
\label{gamma_M_lambda}
\end{center}
\end{figure}

In Fig.~\ref{gamma_M_lambda}, we present the change in the total mass with the adiabatic index for a fixed central rest mass density $\delta_c=1.78266\times10^{16}\,{\rm kg/m^3}$ and for three cosmological constants, namely, $\Lambda=10^{-5}$, $0$ and $-10^{-5}{\rm km}^{-2}$. We also plot the total mass for the null cosmological constant case for the sake of comparison. We take into account stable objects, i.e., static equilibrium configurations with $\omega\geq0$. In the three curves, we note the monotonic growth of the mass with the adiabatic index. 

The influence of the cosmological constant is noticeable in Fig.~\ref{gamma_M_lambda}. When $\Lambda=10^{-5}{\rm km^{-2}}$ ($-10^{-5}{\rm km^{-2}}$), the total mass grows (decreases) with the cosmological constant. This is because $\Lambda$ acts as an effective pressure that aids (hinders) the fluid pressure to support additional mass. The effects of the cosmological constant becomes more noticeable with decreasing $\Gamma$. The softer the matter contained in the spherical object, the greater the effects of the cosmological constant on the structure configurations. In other words, the softer the fluid contained in the object, the lower the cosmological constant value needs to be to affect the physical properties of a star (see \cite{hledik2004}).

\begin{figure}[ht]
\begin{center}
\includegraphics[width=0.49\linewidth]{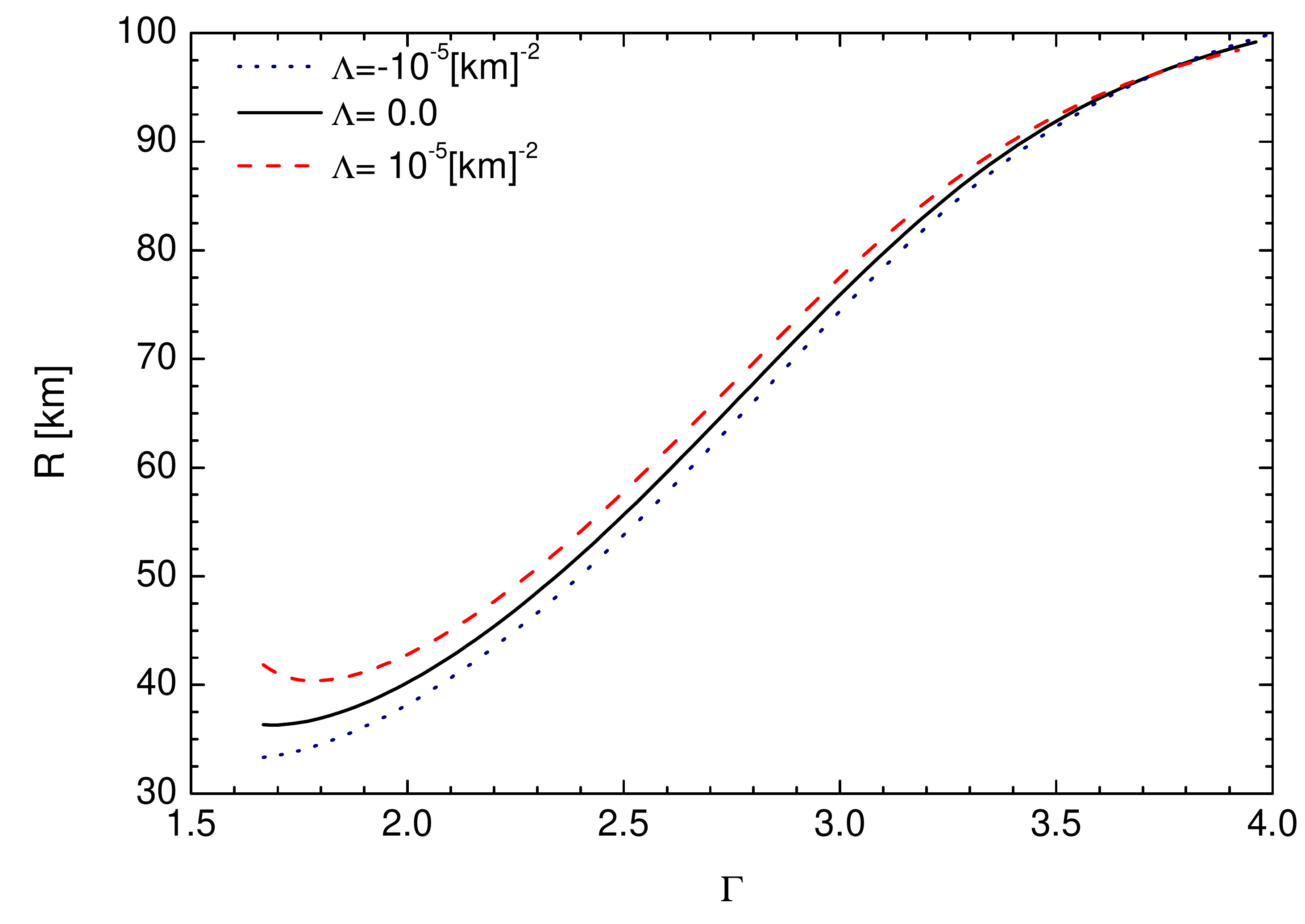}
\caption{Total radius of the object versus the adiabatic index for three different cosmological constant values.}
\label{gamma_R_lambda}
\end{center}
\end{figure}

The total radius as a function of the adiabatic index is plotted in Fig.~\ref{gamma_R_lambda} with $\delta_c=1.78266\times10^{16}\,{\rm kg/m^3}$ for different values of $\Lambda$. As in Fig.~\ref{gamma_M_lambda}, stable stellar configurations against small radial perturbations are considered. When $\Lambda\leq0$, the radius grows monotonically with the adiabatic index. In turn, when $\Lambda>0$, the radius decreases with an increase in $\Gamma$ until $\Gamma\approx1.8$ is reached. After this point, $R$ increases with the adiabatic index.

In Fig.~\ref{gamma_R_lambda}, the influence of the cosmological constant on the  equilibrium configurations can also be seen. When $\Lambda=10^{-5}{\rm km^{-2}}$ ($-10^{-5}{\rm km^{-2}}$), $R$ increases (diminishes) with $\Lambda$. As with the total mass, the effect of the cosmological constant is more evident at relatively low values of $\Gamma$.

\begin{figure}[ht]
\begin{center}
\includegraphics[width=0.49\linewidth]{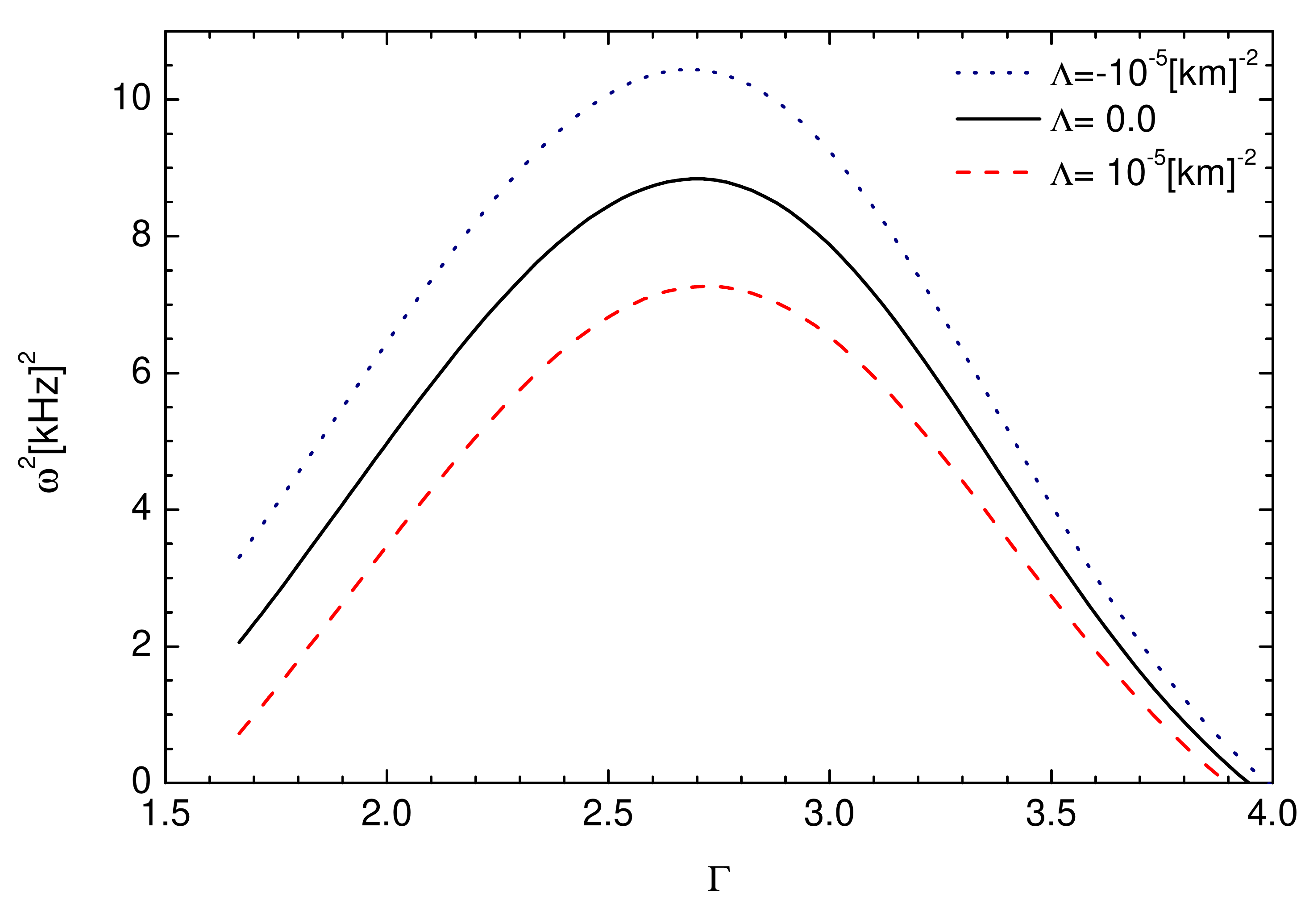}
\caption{Fundamental-mode eigenfrequency of oscillation squared versus the adiabatic index for a few cosmological constant values.}
\label{gamma_w2_lambda}
\end{center}
\end{figure}

The fundamental-mode eigenfrequency of oscillation squared as a function of the adiabatic index is plotted in Fig.~\ref{gamma_w2_lambda} for three different cosmological constant values. In all the curves, we note that $\omega^2$ grows with $\Gamma$ until $\Gamma\approx2.7$. Henceforth, $\omega^2$ decreases with the adiabatic index.

The influence of $\Lambda$ also appears in Fig.~\ref{gamma_w2_lambda}. When $\Lambda=10^{-5}{\rm km^{-2}}$ ($-10^{-5}{\rm km^{-2}}$), $\omega^2$ diminishes (grows) with the cosmological constant. The decline (growth) of the radial stability with the cosmological constant could be inferred by noticing that a positive (negative) value of $\Lambda$ acts as a repulsive (attractive) effective pressure, which is disadvantageous (advantageous) to the radial stability.  The effects of the cosmological constant are more visible for relatively low values of $\Gamma$. As mentioned above, this result could be associated with the hardness of the matter contained in the star; the less hard the matter, the lower the cosmological constant value needs to be to alter the equilibrium configuration.

\section{Conclusions}\label{conclusion}

The cosmological constant has been the foundation for an enormous number of discussions in physics. Although at first glance the importance of the cosmological constant can be conceivable only on cosmological scales, its effects on clusters of galaxies \cite{hameeda/2016,ma/1997}, galaxies \cite{gessner/1992,narlikar/1991,kulchoakrungsun/2018}, gravitational lensing \cite{sereno/2008,turner/1990,biressa/2011,simpson/2010} and even black holes \cite{adams/1999,chirenti/2015,matyjasek/1987} and wormholes \cite{heydarzade/2015,richarte/2013} (in addition to the compact astrophysical objects discussed above) have already been studied.

Within cosmology, beyond the standard model, there is also the possibility that the density of the cosmological constant is increasing with time. This possibility results in phantom energy models, that can yield to a Big Rip fate for the universe \cite{caldwell/2003,vikman/2005,caldwell/2002} (see also References \cite{gonzalez-diaz/2004,briscese/2007,dimopoulos/2018}).

In the present paper, the equilibrium configurations and stability against small radial perturbations of relativistic polytropic objects were analyzed under the Einstein's theory of gravity with the cosmological constant $\Lambda$. For this purpose, we numerically solved the TOV equation and the Chandrasekhar radial oscillation equation, which were altered from their original form to include $\Lambda$. For the fluid contained in the sphere, we considered that the pressure $p$ and the energy density $\rho$ are connected by the form $p=\kappa\delta^{\Gamma}$ with $\delta=\rho-p/(\Gamma-1)$, with $\delta$, $\kappa$ and $\Gamma$ representing the rest mass density, the polytropic constant and the adiabatic index, respectively. Following \cite{raymalheirolemoszanchin,alz-poli-qbh,alz-2eos-qbh}, we considered 
$\kappa=1.47518\times10^{-3}\left(1.78266\times10^{15}{\rm kg/m^3}\right)^{1-\Gamma}$. The effects of the cosmological constant on both the equilibrium and the stability of relativistic polytropic objects were investigated by considering some different central rest mass densities $\delta_c$ and adiabatic indexes $\Gamma$.

The investigation of the present article was initiated by considering a few values of the cosmological constant, some adiabatic index values and varying the central rest mass density. For the relativistic polytropic spheres, an itemized investigation was accomplished by calculating their mass, radius and fundamental-mode eigenfrequency of oscillation. From our results, presented in Section 3, for fixed $\Lambda$ and $\Gamma$, we noted that the maximum mass and the zero eigenfrequency of oscillation are derived with the same central rest mass density. This indicates that in a sequence of equilibrium configurations with fixed $\Lambda$ and $\Gamma$, the necessary and sufficient conditions to recognize regions made by stable and unstable compact objects against small radial perturbations are $dM/d\delta_{c}> 0$ and $dM/d\delta_{c}< 0$, respectively.

Finally, stellar structure configurations were also investigated for fixed $\delta_c$ and $\Lambda$ and several values of $\Gamma$. In this situation, we considered $\delta_c=10\delta_0$, $-10^{-5}\leq\Lambda\leq10^{-5}\,{\rm km^{-2}}$ and $\Gamma$ from $5/3$ to approximately $4$. Indeed, in these ranges of parameters, the influences of the cosmological constant were analyzed. We found that a positive (negative) cosmological constant helps to increase (decrease) the total mass and radius but decreases (increases) the radial stability. These effects are more notable at relatively low values of $\Gamma$, indicating that lower (larger) values of $\Lambda$ are required to influence less (more) compact structures. 

As a possible extension of the present approach, we could, for instance, consider a radially dependent cosmological ``constant.'' The regularity conditions, TOV equation, stability and surface redshift of anisotropic compact stars with $\Lambda(r)$ were analyzed in \cite{hossein/2012}. To do so, the authors assumed a Krori-Barua spacetime metric and the equation of state $p_r=m\rho$, with $p_r$ being the radial pressure and $m>0$ being the equation of state parameter. Their model was shown in some aspects to be similar to boson star models \cite{mielke/2000,sennett/2017,chavanis/2012,gleiser/1988}. 

Another possible extension of this work could come from applying the present approach to a different underlying theory of gravity. In this regard, it is worth mentioning that some alternatives to general relativity are also dependent on a cosmological constant, especially the noncosmologically inspired theories, such as braneworld models \cite{lue/2004,vinet/2004}. In particular, the $f(\mathcal{R},T)$ theory of gravity \cite{harko/2011,msrc/2019,baffou/2017,mam/2016}, where $\mathcal{R}$ is the Ricci scalar and $T$ is the trace of the energy-momentum tensor, can be mapped into a $\Lambda(T)$ gravity model and is welcomed as a foundation for future applications of the present model. 

\begin{acknowledgement}
The authors thank Funda\c{c}\~ao de Amparo \`a Pes\-qui\-sa do Estado de S\~ao Paulo (FAPESP), grant $2013/26258-4$. PHRSM would like to thank FAPESP, grant $2015/08476-0$.
\end{acknowledgement}

\end{document}